\documentclass[aps,pra,twocolumn]{revtex4}%
\usepackage{graphics}
\usepackage{amsmath}
\usepackage{graphicx}
\usepackage{amsfonts}
\usepackage{amssymb}%
\begin{document}

\title{Subwavelength quantum imaging with noisy detectors}
\author{Cosmo Lupo}
\affiliation{Department of Physics \& Astronomy, University of Sheffield, UK}

\begin{abstract}
It has been recently shown that an interferometric measurement may allow for sub-wavelength resolution of incoherent light.
Whereas this holds for noiseless detectors, one could expect that the resolution is in practice limited by the signal-to-noise ratio.
Here I present a qualitative assessment of the ultimate resolution limits that can be achieved using noisy detectors.
My analysis indeed indicates that the signal-to-noise ratio represents a fundamental limit to quantum imaging, and the reduced resolution scales with the square root of the signal-to-noise ratio. For example, a signal-to-ratio of $20 dB$ is needed to resolve one order of magnitude below the wavelength.
\end{abstract}

\maketitle

\section{Introduction}\label{Sec:intro}

The general goal of quantum imaging is to develop methods and techniques that exploit quantum optics to enhance image resolution \cite{Kolobov,Brida,Lugiato,ShapiroBoyd,Perez,Erkmen1,Giovannetti,Unternahrer,Erkmen2,DAngelo,Kok,Costa,Degen,Pirandola}.
An influential work by Tsang, Nair, and Lu \cite{MankeiPRX} has recently put forward a new approach to study quantum imaging using theoretical tools borrowed from quantum 
information theory. This has gathered a certain interest in the quantum optics and quantum information communities, see e.g., \cite{mywork,NairPRL,Sidhu,
Yang1,Ang,Hradil,Backlund,Yu,Napoli,
npj,Dutton,TsangPRA,Paur,Tang,Yang2,Tham,Donohue,Parniak,Stoklasa,Hassett,Zhou,mio2019,Chrostowski,Bonsma-Fisher,R2,Larson1,Tsang000,Z1,Grace,Konrad}.

One of the breakthroughs of Ref.\ \cite{MankeiPRX} was to frame imaging
as a problem of parameter estimation. Given two point-like sources, one faces the task of measuring their transverse separation using an optical imaging system operating in the far field.
Whereas direct detection on the focal plane is limited by the Rayleigh length $\mathsf{x_R} = \lambda R/D$ (where $\lambda$ is the wavelength, $R$ the size of the entrance pupil, and $D$ the distance to the object), Ref.\ \cite{MankeiPRX} showed that this is not the case if a coherent detection scheme is employed.
This is realized by first channelling the light impinging on the focal plane into a multi-port interferometer, and then measuring by photon-detection.
A particular, and optimal, way of realizing this is through spatial mode demultiplexing (SPADE).
In this case the interferometer acts as a mode sorter that decomposes the field in some particular set of normal modes in the transverse field. For example, for a Gaussian point-spread function, these can be chosen as Hermite-Gaussian modes.

Whereas Ref.\ \cite{MankeiPRX} initially assumed noiseless detectors, the impact of detector noise has been addressed only very recently in Ref.\ \cite{Konrad}.
This work has shown that SPADE is in fact limited by the signal-to-noise ratio,  once we depart from the assumption of noiseless detectors and consider dark counts.
However, the fact that SPADE is an optimal detection strategy in the ideal setting does not necessarily imply that it remains optimal with noisy detectors.
Therefore, it is not clear if signal-to-noise ratio is the universal limit to image resolution, or if this is a feature of a particular measurement set up.

In this paper I will address this issue and provide a qualitative assessment of the ultimate resolution of quantum imaging with noisy detectors. 
Following Ref.\ \cite{MankeiPRX}, I will quantify the resolution using the quantum Fisher information for the estimation
of the transverse separation.
The results indicate that signal-to-noise ratio does in fact pose a fundamental limit to quantum imaging.
The effect of detector noise can be described in terms of an effective Rayleigh length that is re-scaled by inverse the square root of the signal-to-noise ratio,
$\mathsf{x_R}' = \mathsf{x_R} / \sqrt{\mathrm{SNR}}$.

The paper will proceed as follows.
Section \ref{Sec:review} reviews the model and methods developed in Ref.\ \cite{mywork}.
Section \ref{Sec:QFI} describes the use of the quantum Fisher information as a theoretical tools to investigate quantum imaging.
Section \ref{Sec:theory} presents a general theory for obtaining the quantum Fisher information for the estimation of the transverse separation between a pair of incoherent sources.
Section \ref{Sec:ideal} applies this theory to the case of noiseless imaging of thermal sources.
A model for noisy imaging is introduced in Section \ref{Sec:darkside}.
This finally allows us to assess the ultimate limits of noisy imaging of thermal sources in in Section \ref{Sec:noise}.
Conclusions are presented in Section \ref{Sec:end}.

\begin{figure}[t]
\centering
\includegraphics[width=0.4\textwidth]{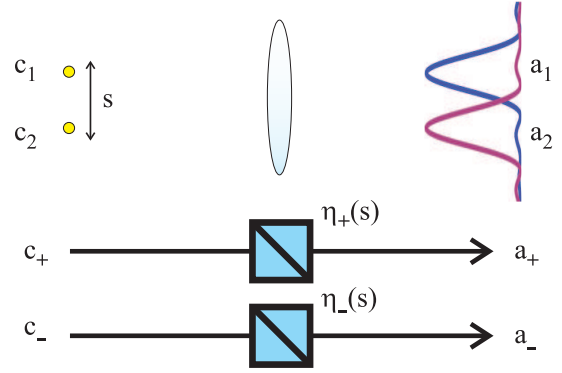}
\caption{
A diffraction-limited system creating an image 
of two point-like sources (top of the figure) is formally equivalent 
to a pair of independent beam-splitters (bottom of the figure), 
whose transmissivities are functions 
of the separation between the sources, 
with $c_\pm = (c_1 \pm c_2)/\sqrt{2}$.}
\label{fig:model}
\end{figure}

\section{The model}\label{Sec:review}

Consider the textbook model of an optical imaging system in the far field as a thin lens with finite aperture, shown in Fig.\ \ref{fig:model} (top).

We denote as $c_1^\dag, c_1$ and $c_2^\dag, c_2$ the creation and annihilation operators associated with two point-like emitters. 
We assume the sources monochromatic and 
separated by a transverse distance $s$. 
They are located in the object plane, orthogonal to the optical axis, at position $-s/2$ and $s/2$.
The optical system transforms these field
operators at the source into corresponding field operators on the image plane, denoted as $a_1^\dag, a_1$ and $a_2^\dag, a_2$.

Assuming, without loss of generality, unit magnification factor, the point-spread function associated to the imaging
system reads \cite{Goodman}
\begin{align}
T(x,y) = \sqrt{\eta} \,  \psi(x-y) \, , 
\end{align}
where $\psi$ is a function on the image plane and $\eta$ is an attenuation factor. The latter accounts for the fact that an optical system in the far field collects only a small fraction of the light coming from the sources.

The image operators $a_1$ and $a_2$ are determined by the point-spread 
function as follows:
\begin{align}
a_1^\dag & = 
\int \, dx \, \psi(x+s/2) \, a_x^\dag \, , \label{defa1} \\
a_2^\dag & = 
\int \, dx \, \psi(x-s/2) \, a_x^\dag \, , \label{defa2}
\end{align}
where $a_x^\dag$ and $a_x$ denote the creation and annihilation operators for 
a field localized at position $x$ on the image screen.

A diffraction-limited optical system does in fact collect
the light coming from the sources and map it into the optical modes
defined by the operators $a_1$, $a_2$ on the image plane.
Therefore, this transformation can be formally represented 
as a beam splitter with transmissivity $\eta$:
\begin{align}
c_1 & \to \sqrt{\eta} \, a_1 + \sqrt{1-\eta} \, v_1 \, , \\
c_2 & \to \sqrt{\eta} \, a_2 + \sqrt{1-\eta} \, v_2 \, ,
\end{align}
where $v_1$, $v_2$ are auxiliary environmental modes that we assumed initially in the vacuum state.

Because of the overlap between the point-spread functions $\psi(x+s/2)$ and $\psi(x-s/2)$, the operators $a_1$ and $a_2$
are not orthogonal each other, i.e., they do not satisfy the canonical commutation relation $[a_1, a_2^\dag] = 0$. 
To orthogonalise them, we define the sum and difference operators 
(see also Refs.\ \cite{Shapiro,mio})
\begin{align}
c_\pm & := \frac{c_1 \pm c_2}{\sqrt{2}} \to \sqrt{\eta_\pm} \, a_\pm + \sqrt{1-\eta_\pm} \, v_\pm \, , \label{c+-} 
\end{align}
where 
\begin{equation}
\eta_\pm = (1 \pm \delta) \eta \, ,
\end{equation}
and we have introduced the operators on the image plane,
\begin{equation}
a_\pm = \frac{a_1 \pm a_2}{\sqrt{2(1 \pm \delta)}} \, ,
\end{equation}
with 
\begin{equation}
\delta = \mathrm{Re} \int dx \, \psi^*(x+s/2) \psi(x-s/2) \, .
\end{equation}
It is easy to check that
operators $a_\pm^\dag$, $a_\pm$ satisfy the canonical commutation relations.

Other parameters of interest are
\begin{align}
\Delta k^2 & := \int dx \left| \frac{\partial\psi(x)}{\partial s} \right|^2 \, , \\
\gamma & := \frac{\partial\delta}{\partial s} \, .
\end{align}

The interpretation of Eq.\ (\ref{c+-}) is that the 
optical modes $c_\pm$ are independently attenuated and mapped into the
modes $a_\pm$, with corresponding effective attenuation factors 
$\eta_\pm = (1\pm\delta)\eta$, see Fig.\ \ref{fig:model} (bottom).
Note that the attenuation factors $\eta_\pm$ depend on the separation
$s$ through the parameter $\delta$.
Therefore, estimating the separation $s$ between two point-like sources
is formally equivalent to estimating the transmissivities of a pair
of independent beam splitter.
This result was obtained and developed in detail in Ref.\ \cite{mywork}.

\section{Imaging as parameter estimation}\label{Sec:QFI}

This section recalls the notion of quantum Fisher information, which here is
used as a theoretical tool to study the ultimate resolution of quantum imaging. 

Consider two emitters separated by a transverse
distance $s$. The light emitted is collected into an optical imaging
system and focused on the image screen (see top of Fig.\ \ref{fig:model}). 
The state of the light focused on the screen
is described by a density matrix $\rho_s$, which is a function of the
transverse separation. 

One can then consider the problem of estimating the parameter $s$
from a measurement of the state $\rho_s$. Given $n$ copies of $\rho_s$
and for any unbiased estimator, the quantum Cram\'{e}r-Rao bound states that \cite{Paris,Jas}
\begin{align}
\Delta s \geq \frac{1}{\sqrt{n \mathrm{QFI}_s(\rho_s)}} \, ,
\end{align}
where $\Delta s$ is the statistical error in the estimation of $s$, and
$\mathrm{QFI}_s(\rho_s)$ is the quantum Fisher information for the estimation
of $s$.

Therefore, the quantum Fisher information quantifies, via the quantum Cram\'{e}r-Rao bound, the ultimate bound in the statistical error for the estimation of $s$. Note that $\mathrm{QFI}_s(\rho_s)$ can be non-zero even when $s$ is zero. 
This is because the statistical error $\Delta s$ can be finite even if the true value of $s$ is zero.

\section{A theory for imaging of incoherent sources}\label{Sec:theory}


Consider a pair of incoherent sources emitting $N$ mean photons each.
The optical imaging system collects, attenuates, and focuses on the image plane the light emitted by the sources.

Given that two identical sources $c_1$, $c_2$ emit incoherent light, this remains incoherent also when expressed in terms of the modes $c_+$, $c_-$. 
As we have discussed in Section \ref{Sec:review}, the optical imaging system is formally equivalent to a passive beam-splitter transformation.
As such, it cannot create coherence in the quantum state.
This implies that the state of the light focused on the image plane, expressed in terms of the normal modes $a_+$, $a_-$, is also incoherent and thermal, and can be described by a two-mode density operator of the form 
\begin{align}\label{state2}
\rho = \rho_+ \otimes \rho_- \, ,
\end{align} 
where
\begin{align}\label{state1}
\rho_\pm = \sum_{n=0}^\infty p_\pm(n) |n\rangle_\pm \langle n|
\end{align} 
is a number-diagonal state, and 
\begin{align}\label{bosonic}
|n\rangle_\pm = (n!)^{-1/2} \left( a_\pm^\dag\right)^n |0\rangle
\end{align} 
denotes a Fock state with $n$ photons.

The problem we need to address here is to compute the quantum Fisher information for the estimation of the parameter $s$ on the state in Eq.\ (\ref{state1}).
Note that such a state depends on $s$ both through the diagonal coefficients $p_\pm(n)$ and through the operators $a_\pm^\dag$ in Eq.\ (\ref{bosonic}).
The contribution to the quantum Fisher information from the coefficients $p_\pm(n)$ is 
\begin{align}
\begin{split}
\langle (\partial_s \log p)^2\rangle 
:=
\sum_{n=0}^\infty p_+(n) \left( \frac{\partial\log p_+(n)}{\partial s} \right)^2 \\
+ p_-(n) \left( \frac{\partial\log p_-(n)}{\partial s} \right)^2 \, ,
\end{split}
\end{align} 
and the contribution from the operators $a_\pm^\dag$ is 
\begin{align}
2 \eta N \left( \Delta k^2 - \frac{\gamma^2}{1-\delta^2} \right) \, .
\end{align} 
For details on how this is obtained see \cite{mywork}.

In conclusions, for states as in Eqs.\ (\ref{state2})-(\ref{state1}), Ref.\ \cite{mywork} obtained an exact expression for the quantum Fisher information:
\begin{align}\label{QFI0}
\mathrm{QFI} = 
\langle (\partial_s \log p)^2\rangle 
+ 2 \eta N \left( \Delta k^2 - \frac{\gamma^2}{1-\delta^2} \right) \, .
\end{align}


\section{Noiseless imaging}\label{Sec:ideal}

For a pair of incoherent thermal sources with $N$ mean photon number we have
\begin{align}\label{prob}
p_\pm(n) =
\frac{1}{M_\pm + 1} \left( \frac{M_\pm}{M_\pm+1} \right)^n \, ,
\end{align} 
with
\begin{align}
M_\pm = \eta_\pm N \, .
\end{align} 

\begin{figure}[t]
\centering
\includegraphics[width=0.4\textwidth]{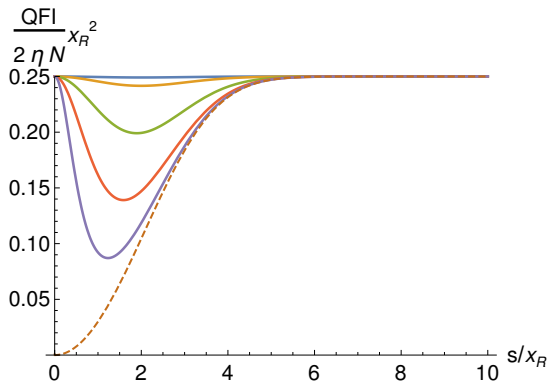}
\caption{
This plot shows the dimensionless quantum Fisher information per photon
detected, $\frac{\mathrm{QFI}}{2 \eta N} \, \mathsf{x_R}^2$, computed from 
Eq.\ (\ref{QFIideal}), versus the dimensionless transverse separation $s/\mathsf{x_R}$.
This is calculated assuming a Gaussian point-spread function as in Eqs.\ (\ref{GaussianPSF0})-(\ref{GaussianPSF3}).
From top to bottom, different lines refer to 
$\eta N = 0.01$, 
$\eta N = 0.1$,
$\eta N = 1$,
$\eta N = 5$,
$\eta N = 20$,
and the classical limit of $N \to \infty$ (dashed line).
The latter is obtained from Eq.\ (\ref{QFIlimit}).}
\label{fig:ideal}
\end{figure}

By applying Eq.\ (\ref{QFI0}) to this example, we obtain an explicit analytical form for the quantum Fisher information for the separation between a pair of thermal sources.
First we obtain
\begin{align}
\begin{split}
\langle (\partial_s \log p)^2\rangle = 
2 \eta N \left[
\frac{\gamma^2}{2(1+\delta)(1+(1+\delta)\eta N)} \right. \\
\left. + \frac{\gamma^2}{2(1-\delta)(1+(1-\delta)\eta N)}
\right] \, .
\end{split}
\end{align} 
Then, putting this expression into Eq.\ (\ref{QFI0}) we finally obtain \cite{mywork}:
\begin{align}\label{QFIideal}
\mathrm{QFI} = 2\eta N \left( \Delta k^2 - 
\frac{\eta N (1+\eta N) \gamma^2}{(1+\eta N)^2 - \delta^2 \eta^2 N^2} \right) \, .
\end{align} 
The quantum Fisher information per photon detected (and re-scaled by the Rayleigh length) is shown in Fig.\ \ref{fig:ideal}.

Figure \ref{fig:ideal} has been obtained
assuming a Gaussian point-spread function, 
\begin{align}\label{GaussianPSF0}
\psi(x) = \sqrt{ \frac{1}{\sqrt{2\pi} \, \mathsf{x_R}} } \, e^{-\frac{x^2}{4 \mathsf{x_R}^2}} \, ,
\end{align}
where $\mathsf{x_R}$ is the Rayleigh length.
This yields
\begin{align}
\delta & = e^{- \frac{s^2}{4\mathsf{x_R}^2} } \, , \label{GaussianPSF1}\\
\Delta k^2 & = \frac{1}{4 \mathsf{x_R}^2} \, , \label{GaussianPSF2}\\
\gamma & = \frac{s^2}{16 \mathsf{x_R}^2} 
\, e^{- \frac{s^2}{4\mathsf{x_R}^2} } \, . \label{GaussianPSF3}
\end{align}

For $s \gg \mathsf{x_R}$ the sources are well separated, and the quantum Fisher information is constant and equal to $2\eta N \Delta k^2$.
The more interesting regime is the sub-Rayleigh region where
$s \lesssim \mathsf{x_R}$.
For $2\eta N \ll 1$ we observe the phenomenon of sub-Rayleigh resolution, this is expressed by the fact that the quantum Fisher information is essentially constant and independent of the value of the separation 
$s$. 
For larger values of $2\eta N$, the quantum Fisher information rapidly decreases for separation of the order of the Rayleigh length.
Eventually, in the classical limit of $N \to \infty$ the quantum Fisher information per photon detected reads
\begin{align}\label{QFIlimit}
\lim_{N\to\infty} \frac{\mathrm{QFI}}{2\eta N} = \Delta k^2 - 
\frac{\gamma^2}{1 - \delta^2}  \, ,
\end{align} 
and is limited by the Rayleigh length. The latter is a known phenomenon dubbed the
{\it Rayleigh curse} \cite{MankeiPRX}.
The same qualitative pattern is observed for 
any arbitrary choice of the point-spread function.

\section{A model for dark counts}\label{Sec:darkside}

Section \ref{Sec:review} has established that the optical imaging system is described, in the Heisenberg picture, by the map of Eq.\ (\ref{c+-}):
\begin{align}\label{c+-2}
c_\pm & := \frac{c_1 \pm c_2}{\sqrt{2}} \to \sqrt{\eta_\pm} \, a_\pm + \sqrt{1-\eta_\pm} \, v_\pm \, .  
\end{align}
This map transforms the bosonic operators $c_\pm$, which describe the sources, into the operators $a_\pm$, which describe the field on the image screen.
This is a linear transformation formally equivalent to a beam-splitter mixing a signal with the vacuum.

To model dark counts in noisy detectors, I will modify Eq.\ (\ref{c+-2}).
I assume dark counts follow a thermal distribution characterised by an effective mean photon number $\epsilon$.
Given the stochastic character of dark counts, I will model them by introducing a pair of additive Gaussian random variables in Eq.\ (\ref{c+-2}), yielding
\begin{align}\label{c+-3}
c_\pm & := \frac{c_1 \pm c_2}{\sqrt{2}} \to \sqrt{\eta_\pm} \, a_\pm + \sqrt{1-\eta_\pm} \, v_\pm + \xi_\pm \, , 
\end{align}
where $\xi_+$ and $\xi_-$ are independent Gaussian random variables with zero mean and variance $\epsilon$.

With this modification to the map describing the optical imaging system, the state of the light impinging on the image screen is still of the form in Eqs.\ (\ref{state2})-(\ref{state1}), but with 
\begin{align}\label{pnoise}
p_\pm(n) =
\frac{1}{\eta_\pm N + \epsilon + 1} \left( \frac{\eta_\pm N + \epsilon}{\eta_\pm N + \epsilon+1} \right)^n \, .
\end{align} 

Note that with this model the dark counts, which are a feature of the detectors, are described as a feature of the state that is to be measured.

\section{Quantum Fisher information of noisy imaging}\label{Sec:noise}

In Ref.\ \cite{Konrad}, Let et al.\ presented an analysis of the ultimate resolution that can be attained using SPADE with non-ideal detectors affected by dark counts. 
Whereas SPADE is known to saturate the quantum Cram\'{e}r-Rao bound with ideal, noiseless, detectors, and to yield super-resolution, Ref.\ \cite{Konrad} showed that resolution is in fact limited by the signal-to-noise ratio.
They observed that the Fisher information decreases substantially when the separation between the source is of the order of $\mathrm{SNR}^{-1/2}$

Note that, though SPADE is optimal with ideal detectors, there is no guarantee that is remains such in the presence of noise.
To see if this is the case we would need to compute the quantum Fisher information for noisy detectors, which is an obvious contradiction. 
In fact, the quantum Fisher information is, by definition, an universal bound that is independent of any specific measurement.

To sidestep this problem and obtain a qualitative but physically sound result, here I have introduced in Section \ref{Sec:darkside} a model for dark counts described as thermal background radiation.
Within this model, the field on the image screen is described by a state of the form
\begin{align}\label{state2_b}
\rho = \rho_+ \otimes \rho_- \, ,
\end{align} 
where
\begin{align}\label{state1_b}
\rho_\pm = \sum_{n=0}^\infty p_\pm(n) |n\rangle_\pm \langle n| \, ,
\end{align}
and $p_\pm(n)$ as in Eq.\ (\ref{pnoise}).

I can therefore apply the general theory of Section \ref{Sec:theory} and obtain, putting Eq.\ (\ref{pnoise}) in Eq.\ (\ref{QFI0}) :
\begin{align}
& \mathrm{QFI} = 2 \eta N \left( \Delta k^2 - \frac{\gamma^2}{1-\delta^2} \right) \nonumber \\
& + 2\eta N \gamma^2 
\frac{\left(\frac{\epsilon}{\eta N}+1\right)(\eta N + \epsilon + 1) + \delta^2 \eta N}{\left(\left(\frac{\epsilon}{\eta N}+1\right)^2-\delta^2\right)\left( (\eta N + \epsilon +1)^2 - \delta^2 \eta^2 N^2 \right)}
\, . \label{QFInoisy}
\end{align}

%



\begin{figure}[t]
\centering
\includegraphics[width=0.4\textwidth]{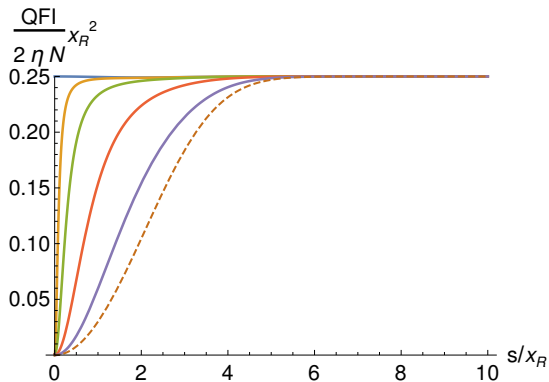}
\caption{
For noisy detectors, this shows the dimensionless quantum Fisher 
information per signal photon
detected, $\frac{\mathrm{QFI}}{2 \eta N} \, \mathsf{x_R}^2$, computed from 
Eq.\ (\ref{QFInoisy}), versus the dimensionless transverse separation $s/\mathsf{x_R}$.
This is calculated putting $\eta N = 0.01$, and for a Gaussian point-spread function as in Eqs.\ (\ref{GaussianPSF0})-(\ref{GaussianPSF3}).
Different lines refer to different values of the signal-to-noise ratio, $\mathrm{SNR} = \eta N/\epsilon$.
From top to bottom, 
the noiseless limit, 
$\mathrm{SNR} = 10^3$,
$\mathrm{SNR} = 10^2$,
$\mathrm{SNR} = 10$,
$\mathrm{SNR} = 1$,
and in the limit of $\mathrm{SNR} \to \infty$ (in dashed line).
The latter is identical to the dashed line in Fig.\ \ref{fig:ideal}.}
\label{fig:noisy}
\end{figure}

This quantity is plotted in Fig.\ \ref{fig:noisy} for different values of the
signal-to-noise ratio, $\mathrm{SNR} = \eta N/\epsilon$, for a Gaussian point-spread function (see Eqs.\ (\ref{GaussianPSF0})-(\ref{GaussianPSF3})).
For $s \gg \mathsf{x_R}$ the sources are well separated and the quantum Fisher information is constant and equal to $ 2 \eta N \Delta k^2$, as in the ideal set up.

For sub-Rayleigh separation, it is the signal-to-noise ratio that
determines the ultimate precision bound. Following Ref.\ \cite{Konrad}, define $s_{1/2}$ as the value of $s$ such that $\mathrm{QFI} = \eta N \Delta k^2$, i.e.\ the separation at which the quantum Fisher information attains one half of its maximum value. 
Then we obtain, in the interesting regime of $\eta N \ll 1$, and for low detector noise, $\mathrm{SNR} \gg 1$, the following expression for $s_{1/2}$ at the lowest order in $s_{1/2}/\mathsf{x_R}$ (note that in this regime $s_{1/2}/\mathsf{x_R} \ll 1$), 
\begin{align}
s_{1/2} \simeq \frac{ 8 \mathsf{x_R} }{\sqrt{\mathrm{SNR}}} \, .
\end{align}
This expression has been obtained from Eq.\ (\ref{QFInoisy}) assuming a Gaussian point-spread function. This confirms the scaling $s_{1/2} \simeq \mathrm{SNR}^{-1/2}$ observed in Ref.\ \cite{Konrad} for SPADE.

For $\mathrm{SNR} \ll 1$ we instead obtain
\begin{align}\label{QFIlimit1}
\lim_{\mathrm{SNR} \to 0} \frac{\mathrm{QFI}}{2\eta N} = \Delta k^2 - 
\frac{\gamma^2}{1 - \delta^2}  \, ,
\end{align} 
which is identical to the classical limit in Eq.\ (\ref{QFIlimit}).
Note that this limit is essentially already achieved for $\mathrm{SNR} \simeq 0.1$.
This implies that, by increasing the mean number of thermal
photons, the resolution eventually becomes Rayleigh-limited, irrespective of
whether these photons originated form the signal or from the noise.

\section{Conclusions}\label{Sec:end}

I have presented an analysis of the ultimate resolution
of quantum imaging that one can attain using noisy detectors affected by
dark counts.
This is done within the approach put forward by Tsang and collaborators in
Ref.\ \cite{MankeiPRX} for the case of two incoherent sources of thermal
light.

As first noted in Ref.\ \cite{MankeiPRX}, for weak sources one observes a phenomenon of sub-Rayleigh resolution. This is expressed by the fact that the quantum Fisher information for the estimation of the transverse separation is constant and independent of the value of the separation, also if this is
far below the Rayleigh length. 
This is in contrast with the semi-classical limit, obtained by increasing the mean photon number, where the quantum Fisher information rapidly goes to zero when the transverse separation is comparable with the Rayleigh length.

I have obtained a closed formula for the quantum Fisher information in the presence of thermal background noise, which models detectors dark counts.
The results presented here confirm the findings that Len et al.\ have presented in Ref.\ \cite{Konrad} for specific types of measurement.
The ambition of my work is to assess the impact of noise on the resolution of quantum imaging without assuming a specific measurement strategy.
In order to do this, I have used the quantum Fisher information as a theoretical tool and modeled the noise from the detectors as a thermal background.

Note that this way of computing the quantum Fisher information, though physically sound, cannot be exact. 
This is because dark counts are a feature of the measurement device, whereas the quantum Fisher information is, by definition, independent of the measurement apparatus.
For this reason, it would be meaningless to search for the optimal measurement that saturates the quantum Fisher information of Eq.\ (\ref{QFInoisy}).
If such an optimal measurement existed, it would be noiseless.
This would be a contradiction as the Eq.\ (\ref{QFInoisy}) is intended to describe noisy detectors.


In conclusion, my analysis is physically sound but, by construction, it cannot be completely self-consistent.
More work is needed to develop the ideas presented here into a complete and self-consistent theory.
Nevertheless, this work suggests that the detector noise degrades the super-resolution phenomenon and introduces a new resolution cutoff. 
For high signal-to-noise (low noisy), this cutoff happens when the transverse separation is of the order of $\mathsf{x_R} / \sqrt{\mathrm{SNR}}$, where
$\mathsf{x_R}$ is the Rayleigh length and $\mathrm{SNR}$ is the signal-to-noise ratio.
This is consistent with the results of Ref.\ \cite{Konrad}, which have been obtained with a different method.
Otherwise, for small signal-to-noise ratio (high noise), one recovers the semi-classical limit and the Rayleigh curse.

\subsection*{Acknowledgements}


This  work  was  supported  by  the  UK  Engineering and Physical  Sciences  Research  Council  (EPSRC)  through  UK Quantum  Technology  Hub  for  Quantum  Communications Technologies, Grant No. EP/M013472/1. 
I acknowledge comments from Z. Huang and Y. L. Len on an early version of the manuscript.

\end{document}